\def\rfr#1{eq. (\ref{#1})}

\def\bar{\begin{eqnarray}}
\def\ear{\end{eqnarray}}
\def\bb{\bibitem}
\def\eqi{\begin{equation}}
\def\eqf{\end{equation}}
\def\eqia{\begin{eqnarray}}
\def\eqfa{\end{eqnarray}}
\def\rp#1#2{{#1\over#2}}

\def\lb#1{\label{#1}}

\def\oc2{$\mathcal{O}(c^{-2})$}

\documentclass[11pt]{article}
\usepackage{amsmath,amsthm,amscd,amssymb}
\usepackage{latexsym}
\usepackage{graphicx,epsfig}

\begin{document}

\noindent{\bf \LARGE{Test of the Pioneer anomaly with the Voyager
2 radio-ranging distance measurements to Uranus and Neptune}}
\\
\\
\\
{Lorenzo Iorio}\\
{\it Viale Unit$\grave{a}$ di Italia 68, 70125\\Bari, Italy
\\tel./fax 0039 080 5443144
\\e-mail: lorenzo.iorio@libero.it}

\begin{abstract}
In this paper we test the hypothesis that the Pioneer anomaly can
be of gravitational origin by comparing the predicted
model-independent shifts $\Delta a/a$ for the semimajor axis of
Uranus and Neptune with the Voyager 2 radio-technical distance
measurements performed at JPL-NASA. As in the case of other tests
based on different methods and data sets (secular perihelion
advance, right ascension/declination residuals over about one
century), the orbits of the investigated planets are not affected
by any anomalous acceleration like that experienced by the Pioneer
10/11 spacecraft.
\end{abstract}



\section{Introduction}
The Pioneer anomaly (Anderson et al. 1998; 2002) consists of an
unexpected, almost constant and uniform acceleration directed
towards the Sun \eqi A_{\rm Pio}=(8.74\pm 1.33)\times 10^{-10} \
{\rm m\ s}^{-2}\lb{pioa}\eqf detected in the data of both the
spacecraft Pioneer 10 (launched in March 1972) and Pioneer 11
(launched in April 1973) after they passed the threshold of 20
Astronomical Units (AU; 1 AU is slightly less than the average
Earth-Sun distance and amounts to about 150 millions kilometers).
Latest communications with the Pioneer spacecraft, confirming the
persistence of such an anomalous feature, occurred when they
reached 40 AU (Pioneer 11) and 70 AU (Pioneer 10).

If the Pioneer anomaly is of gravitational origin, it must then
fulfil the equivalence principle, which is presently tested at a
$10^{-12}$ level (Will 2006) and lies at the foundations of the
currently accepted theories of gravity. In its weak form, it
states that different bodies fall  with the same accelerations in
a given external gravitational field. As a consequence, an
extra-gravitational acceleration like $A_{\rm Pio}$ should also
affect the motion of any other object moving, at least, in the
region in which the Pioneer anomaly manifested itself.

In this context, many models have been proposed in order to find
some possible gravitational explanation of the anomalous
acceleration experienced by the Pioneer 10/11 spacecraft. E.g.,
Jaekel and Reynaud (2005) put forth a metric linear extension of
general relativity which yields an acceleration only affecting the
radial component of the velocity of a test particle. Brownstein
and Moffat (2006) used an explicit, analytical model fitted  to
all the presently available Pioneer 10/11 data points.

The availability of
\begin{itemize}
  \item
  The latest observational determinations of the secular, i.e.
averaged over one orbital revolution, extra-advances of perihelia
$\dot\varpi$ of the inner (Pitjeva 2005a) and of some of the outer
(Pitjeva 2006a) planets of the Solar System
  \item The residuals of the direct observables $\alpha\cos\delta$ and
$\delta$, where $\alpha$ and $\delta$ are the right ascension and
the declination, respectively, for the gaseous giant planets and
Pluto (Pitjeva 2005b)
\end{itemize}
processed at the Institute of Applied Astronomy, Russian Academy
of Sciences (IAA, RAS)
 has recently allowed to

\begin{itemize}
 \item  Perform clean and
unambiguous tests of the possibility that the acceleration of
\rfr{pioa} can also affect the planetary motions in the
far\footnote{Sanders (2006) disproves the possibility that $A_{\rm
Pio}$ is also present in the inner regions of the Solar System by
using older data for the perihelion advances of Mercury and
Icarus. It is straightforward to use the more recent and accurate
data for all the inner planets (Pitjeva 2005a) to put much tighter
constraints on any possible anomalous extra-acceleration in such a
region of the Solar System. } regions of the Solar System (Iorio
2006a; Iorio and Giudice 2006)

\item Dismiss the previously cited mechanisms for the anomalous
Pioneer behavior (Iorio 2006b; 2006c)
\end{itemize}

In this letter we perform a further, independent test of the
hypothesis that the Pioneer anomaly can be of gravitational origin
by exploiting certain short-period, i.e. not averaged over one
revolution, features of the semimajor axes $a$ of Uranus and
Neptune and the radar-ranging distance measurements to them
performed at Jet Propulsion Laboratory (JPL), NASA, during their
encounters with the Voyager 2 spacecraft (Anderson et al. 1995).

The outcome of such a test is consistent with the other ones based
on $\dot\varpi$ (Iorio 2006a; 2006c) and $\alpha\cos\delta-\delta$
(Iorio and Giudice 2006): an acceleration like that of \rfr{pioa}
does not affect the motion of Uranus and Neptune.


\section{The effect of a Pioneer-like acceleration on the semimajor axis and comparison with the observations}
In (Iorio and Giudice 2006) there are the analytical expressions
of the short-period shifts induced on the Keplerian orbital
elements by a radial, constant perturbing acceleration $A_{r}$,
whatever its physical origin may be. For the semimajor axis we
have \eqi \rp{\Delta a}{a}=-\rp{2eA_{r }a^2}{GM}(\cos E-\cos E_0
)=-\rp{2e}{\sqrt{1-e^2}}\rp{A_r}{\left\langle A_{\rm
N}\right\rangle}(\cos E-\cos E_0),\lb{pippo}\eqf where $GM$ is the
Kepler's constant, $e$ is the orbital eccentricity,
\eqi\left\langle A_{\rm
N}\right\rangle=GM\left<\rp{1}{r^2}\right>=\rp{GM}{a^2\sqrt{1-e^2}}\eqf
is the Newtonian acceleration averaged over one orbital period and
$E$ is the eccentric anomaly which can be expressed in terms of
the mean anomaly $\mathcal{M}$ as (Roy 2005)\eqi E \sim
\mathcal{M}+\left(e-\rp{e^3}{8}\right)\sin
\mathcal{M}+\rp{e^2}{2}\sin 2\mathcal{M}+\rp{3}{8}e^3\sin
3\mathcal{M}.\eqf The reference epoch is customarily assumed to be
J2000, i.e. JD=2451545.0 in Julian date. From \rfr{pippo} it can
be noted that, whatever the eccentricity of the orbit is,
  \eqi\left\langle\rp{\Delta a}{a}\right\rangle=0,\eqf so
  that $\Delta a/a$ cannot tell us anything about the impact of an acceleration
  like $A_{\rm Pio}$ for those planets for which data sets covering
  at least one full orbital revolution  exist. To date, only Neptune ($P=164$ yr) and Pluto ($P=248$ yr) have not yet described a full orbit since
  modern astronomical observations became available after the first decade of 1900.
Incidentally, let us note that\footnote{In the circular orbit
limit, Anderson et al. (1998; 2002) use the erroneous formula
$\Delta a/a=-A_r/A_{\rm N}$ and apply it to Mars and the Earth to
show that an extra-acceleration like $A_{\rm Pio}$ cannot exist in
that regions of the Solar System.}, according to \rfr{pippo},
$\Delta a/a=0$ for $e=0$.

  The situation is different for Neptune since no secular effects can yet be measured for it.
   Thus, let us use \rfr{pippo} and
  \rfr{pioa} for $A_r$
  getting
  \eqi \left.\rp{\Delta a}{a}\right|_{\rm Nep}=(-2.2882\pm 0.3482)\times 10^{-6}(\cos E-\cos E_0).\lb{effetto}\eqf
  The predicted effect of \rfr{effetto} can be compared with the
  latest available observational determinations. Pitjeva (2005b) used
  only
  optical data (Table 3 of (Pitjeva 2005b)) for the outer planets (apart from Jupiter) obtaining a
  formal, statistical error  $\delta a=478532$ m
  for the Neptune's semimajor axis (Table 4 of (Pitjeva 2005b)) at JD=2448000.5 epoch (Pitjeva 2006b).
  By re-scaling it by $10-30$
  times in order to get realistic uncertainty we
  get\eqi \left.\rp{\delta a}{a}\right|_{\rm Nep}^{(\rm optical)}=(1-3)\times
  10^{-6}.\eqf It must be compared with \rfr{effetto} at\footnote{For Neptune $E_0=128.571$ deg at JD=2451545.0.} JD=2448000.5 ($E=107.423$ deg)
\eqi\left.\rp{\Delta a}{a}\right|_{\rm Nep}({\rm
JD}=2448000.5)=(-0.7413\pm 0.1128)\times 10^{-6}.\eqf Such an
effect would be too small to be detected.

  In (Anderson et al. 1995) the radio-technical data of the Voyager 2 encounter with Neptune
  were
  used yielding a unique ranging measurement of $a$ (Julian Date
  JD=2447763.67); \rfr{effetto}, evaluated at such epoch ($E=106.012$ deg), predicts
  \eqi\left.\rp{\Delta a}{a}\right|_{\rm Nep}({\rm JD}=2447763.67)=(-0.7954\pm 0.1210)\times 10^{-6}.\lb{predi}\eqf
  By assuming for $\Delta a$ the residuals with respect to the DE200 JPL
  ephemerides used in Table 1 of (Anderson et al. 1995), i.e. $8224.0\pm 1$ km, one
  gets
  \eqi\left.\rp{\Delta a}{a}\right|_{\rm Nep}^{(\rm ranging)}=(1.8282\pm 0.0002)\times
  10^{-6}.\eqf
  This clearly rules out the prediction of \rfr{predi}.

  The same analysis can also be  repeated for Uranus ($P=84.07$ yr) for which no
  modern data covering a full orbital revolution were available at
  the time of the Anderson et al. (1995) work; as for Neptune, one radar-ranging distance measurement is available
  from the Voyager 2 flyby with Uranus (JD=2446455.25). The prediction of
  \rfr{pippo}, with \rfr{pioa} for $A_r$, for the flyby epoch\footnote{For Uranus $E_0=70.587$ deg at JD=2451545.0.}
  ($E=8.860$ deg) is
\eqi\left.\rp{\Delta a}{a}\right|_{\rm Ura}({\rm
JD}=2446455.25)=(-3.3576\pm 0.5109)\times 10^{-6}.\lb{prediu}\eqf
Table 1 of (Anderson et al. 1995) yields for the DE200 residuals
of the Uranus' semimajor axis $\Delta a=147.3\pm 1$ km, so that
\eqi\left.\rp{\Delta a}{a}\right|^{(\rm ranging)}_{\rm Ura}({\rm
JD}=2446455.25)=(0.0513 \pm 0.0003)\times 10^{-6}.\lb{measu}\eqf
Also in this case, the effect which would be induced by $A_{\rm
Pio}$ on $\Delta a/a$ is absent.

It maybe interesting to note that the paper by Anderson et al.
(1995) has been used as a basis for other tests with the outer
planets using different methods. E.g., Wright (2003) and Sanders
(2006) adopt the third Kepler's law. Basically, the line of
reasoning is as follows. In the circular orbit limit, let us
write, in general, $P=2\pi a/v$; in particular, the third Kepler
law states that $P=2\pi\sqrt{a^3/K_p}$, where $K_p=GM_{\odot}$. If
we assume that $K_p$ may vary by $\Delta K_p$ for some
reasons\footnote{E.g. due to dark matter (Anderson et al. 1995).}
inducing a change in the orbital speed, then $\Delta
v/v=(1/2)\Delta K_p/K_p$. In general, for an additional radial
acceleration acting upon a test particle in circular orbit $\Delta
A$, $\Delta A/A=2\Delta v/v$: thus, we have \eqi\rp{\Delta
K_p}{K_p}=\rp{\Delta A}{A}.\eqf Now, a measurement of the planet's
velocity is needed to get $\Delta K_p/K_p$ (or, equivalently,
$\Delta A/A$): since $v=na$, where $n$ is the orbital frequency,
this requires a measurement of both $a$ and $n$, while in our case
we only use $a$. Moreover, the measurement of the orbital
frequency pose problems for such planets which have not yet
completed a full orbital revolution, as it was the case for Uranus
and Neptune at the time of the analysis by Anderson et al. (1995).
For Neptune, according to the last row of Table 2 of (Anderson et
al. 1995), $\Delta K_p/K_p^{\rm meas}=(-2.0\pm 1.8)\times
10^{-6}$, while $A_{\rm Pio}/A_{\rm N}=(-133.2\pm 20.3)\times
10^{-6}$. As can be noted, also in this case, the answer is
negative  but the accuracy is far worse than in our test.

\section{Discussion and conclusions}
In this paper we have used the NASA-JPL radio-technical ranging
measurements to Uranus and Neptune performed during the  Voyager 2
flybies (Anderson et al. 1995) in order to make a
model-independent test of the hypothesis that an anomalous
acceleration of gravitational origin like that detected in the
data of the Pioneer 10/11 spacecraft may also affect the orbital
motion of such planets. The answer is neatly negative, as in
previous tests involving the perihelion secular advance of Uranus
(Iorio 2006a; 2006c) and the right ascension/declination residuals
of Uranus, Neptune and Pluto over about one century (Iorio and
Giudice 2006).

Thus, in regard to the  celestial bodies lying at the edge of the
region in which the Pioneer anomaly manifested itself ($\sim
20-70$ AU), or entirely residing in it, the present-day situation
can be summarized  as follows
\begin{itemize}
  \item Uranus ($a=19.19$ AU). 3 model-independent tests
\begin{itemize}
  \item Secular advance of perihelion (almost one century of  optical data processed at IAA, RAS): negative
  \item Right ascension/declination residuals (almost one century of optical data processed at IAA, RAS):
  negative
  \item Short-period semimajor axis shift (1 radar-ranging measurement at epoch JD=2446455.25 by JPL,
  NASA): negative
\end{itemize}
  \item Neptune ($a=30.08$ AU). 2 independent tests
\begin{itemize}
  \item Right ascension/declination residuals (almost one century of optical data processed at IAA, RAS):
  negative
  \item Short-period semimajor axis shift (1 radar-ranging measurement at epoch JD=2447763.67 by JPL,
  NASA): negative
\end{itemize}

  \item Pluto ($a=39.48$ AU). 1 test
\begin{itemize}
  \item Right ascension/declination residuals (almost one century of optical data processed at IAA, RAS):
  negative
\end{itemize}

\end{itemize}
In all such tests the observationally determined
quantities$-$obtained at JPL and IAA independently and without
having the Pioneer anomaly in mind at all$-$have been compared to
unambiguous theoretical predictions based on the effects induced
by a radial, constant and uniform acceleration with the same
magnitude of that experienced by Pioneer 10/11, without making any
assumptions about its physical origin.

In addition, we may also consider the perihelion-based negative
tests for Jupiter ($a=5.20$ AU) and Saturn ($a=9.53$ AU) (Iorio
2006c), based on the model by Brownstein and Moffat (2006) fitted
to all the presently available data points of Pioneer 10/11.

In conclusion, it seems more and more difficult to accept the
possibility that some modifications of the current laws of
Newton-Einstein gravity may be the cause of the Pioneer anomaly.

\section*{Acknowledgements}
I thank Edward L. (Ned) Wright for useful correspondence about the
radar-ranging distance measurements of Uranus and Neptune and for
the reference to his webpage.


\end{document}